# Power-law distributions based on exponential distributions: Latent scaling, spurious Zipf's law, and fractal rabbits

Yanguang Chen

(Department of Geography, College of Urban and Environmental Sciences, Peking University, 100871, Beijing, China. Email: chenyg@pku.edu.cn)

**Abstract:** The different between the inverse power function and the negative exponential function is significant. The former suggests a complex distribution, while the latter indicates a simple distribution. However, the association of the power-law distribution with the exponential distribution has been seldom researched. Using mathematical derivation and numerical experiments, I reveal that a power-law distribution can be created through averaging an exponential distribution. For the distributions defined in a 1-dimension space, the scaling exponent is 1; while for those defined in a 2-dimension space, the scaling exponent is 2. The findings of this study are as follows. First, the exponential distributions suggest a hidden scaling, but the scaling exponents suggest a Euclidean dimension. Second, special power-law distributions can be derived from exponential distributions, but they differ from the typical power-law distribution. Third, it is the real power-law distribution that can be related with fractal dimension. This study discloses the inherent relationship between simplicity and complexity. In practice, maybe the result presented in this paper can be employed to distinguish the real power laws from spurious power laws (e.g., the fake Zipf distribution).

# 1 Introduction

Many scientists have been attracted by power laws, especially, the inverse power laws of nature, because the scaling invariance of power-law distributions suggests complexity. Only a few scholars take interest in the relationship between power laws and exponential laws. Previous studies showed that a power law can be decomposed into two exponential laws based on hierarchical structure (Chen, 2012; Gutenberg and Richter, 1954; Horton, 1945; Philips, 1999; Schumm, 1956; Strahler, 1952; Turcotte, 1997). In fact, a fractal, or a hierarchy with cascade structure, can be described with both power functions and exponential functions (Chen and Zhou, 2003). Fractals open a good window for examining the relationships between power laws and exponential laws. However, where fractals are concerned, exponential laws and power laws are applied to different directions (Chen, 2012). Exponential functions are applied to the distribution of different levels (a longitudinal description), while power functions are applied to the relation of different measurements (a transverse description). So far, we have limited knowledge about the exponential distributions and power-law distributions which are defined in the same direction.

Despite the differences between the exponential functions and the power functions, they can be employed to describe the same type of distribution or form in urban studies. For example, for inter-urban systems of cities, both the exponential functions and power functions can be applied to the rank-size distribution of settlements; for intra-urban systems, both the two functions can be used to describe urban density (Batty and Longley, 1994; Cadwallader, 1996; Chen and Feng, 2012; Longley et al, 1991; Longley and Mesev, 1997; McDonald, 1989; Makse et al, 1995; Makse et al, 1998; Parr, 1985a; Parr, 1985b; Wang and Zhou, 1999; Zielinski, 1979). The rank-size distributions of urban settlements that follow power laws can be described with Zipf's law or Pareto's law (Carroll, 1982; Gabaix and Ioannides, 2004). However, there seems to be more than one power law for Zipfian distribution (Cristelli et al, 2012). On the other hand, the rank-size distribution of rural settlements may take on exponential laws, which cannot be described with Zipf's law or Pareto's law (Grossman and Sonis, 1989; Sonis and Grossman, 1984). Urban population density always follows an exponential law and can be described with Clark's model (Clark, 1951), but traffic network density within a urban region always follows an inverse power law and can be described with Smeed's model (Smeed, 1963). These phenomena suggest that there exist some inherent relationships between the power-law distributions and the exponential distributions.

In fact, the exponential law and the power law are two basic mathematical laws followed by many observations of the ubiquitous empirical patterns in physical and social systems. The significance of the studies on the relationships between exponential laws and power laws goes beyond geography. Based on the empirical results of urban studies, this paper is devoted to exploring the power-law distributions resulted from exponential distributions. In Section 2, two power-law distributions will be created by averaging exponential distributions defined in 1-dimension and 2-dimension space, respectively. In Section 3, mathematical experiments will be employed to validate the theoretical inferences on the link between exponential law and power law. In Section 4, several questions will be discussed. Finally, the article will be concluded by summarizing the main points of this work. The principal contributions of this paper to academic studies are as follows. First, a new link of power law to exponential law is found. This is revealing for our understanding the relationships between simplicity and complexity. Second, the hidden scaling of exponential distribution is brought to light. The latent scaling is useful for exploring the scale-free distributions of complex systems. Third, a spurious Zipf distribution is proposed. This distribution is helpful for understanding the real Zipf distribution.

## 2 Mathematical derivations

### 2.1 The power law based on exponential distribution in 1-D space

The phenomena of exponential distributions are everywhere in the real world. For example, if the size distribution of cities in a region follows Zipf's law, the cities can be organized into a hierarchy with cascade structure, and the average sizes of different levels follow the negative exponential law (Chen, 2012; Chen and Zhou, 2003). The exponential distribution defined in 1-dimension space can be expressed as

$$y = y_0 e^{-r/r_0}, \tag{1}$$

where $r$ is an independent variable (argument) associated with time or spatial distance, $y$ denotes the corresponding dependent variable indicative of some kind of frequency, the proportionality parameter $y_0$ suggests an initial value of $y$, the scale parameter $r_0$ suggests a characteristic length of this distribution. Integrating $y$ over $r$ yields a cumulative distribution function such as

$$S(r) = y_0 \int_0^r e^{-x/x_0} dx = S_0 (1 - e^{-r/r_0}), \tag{2}$$

where $S(r)$ refers to the cumulative exponential distribution, the coefficient

$$S_0 = r_0 y_0 \tag{3}$$

indicates a capacity of the cumulative result. That is, if $r \to \infty$, then $S(r)=S_0$. Equation (3) is based on continuous variables. For the discrete variables, the characteristic parameter can be expressed as

$$r_0 = \frac{S_0 \Delta r}{y_0} \xrightarrow{\Delta r=1} \frac{S_0}{y_0}. \tag{4}$$

An interesting discovery is that, under the condition of $r > r_0$, the average distribution of the exponents leads to an inverse power-law distribution. Based on equation (2), the average distribution of equation (1) can be defined as

$$F(r) = \frac{S_0}{r}[1 - \exp(-\frac{r}{r_0})], \tag{5}$$

where $r > 0$, and $F(r)$ refers to an average distribution of the exponential distribution defined in a 1-dimensin space. It can be demonstrated that equation (5) possesses a hidden scaling. The scaling transform is as follows

$$F(\lambda r) = \frac{S_0}{\lambda r}[1 - \exp(-\frac{\lambda r}{r_0})] = \frac{1}{\lambda}\frac{S_0}{r}[1 - \exp(-\frac{r}{r_0})] = \lambda^{-1} F(r), \tag{6}$$

which is in fact a functional equation. The key is that, if $r > r_0$ and the value of $r_0$ is small enough, the cumulative distribution satisfies the following approximate relation

$$1 - e^{-\lambda r / r_0} \xrightarrow{r > r_0} 1 - e^{-r/r_0}. \tag{7}$$

This can be validated by mathematical experiments conducted in Microsoft Excel, and partial results will be shown in Section 3. The solution to the equation (6) is a special inverse power function such as

$$F(r) \propto r^{-1}, \tag{8}$$

which is just the mathematical form of Zipf' law, rank-size rule, $1/f$ noise, and so on.

## 2.2 The power law based on exponential distribution in 2-D space

The exponential distribution of a 2-dimension space can be defined in a 1-dimension space using the idea of statistical average. The reduction of spatial dimension from 2 to 1 makes the mathematical treatment become simple. Clark's exponential model of urban population density is just defined in a 1-dimension space for the 2-distribution of population (Chen and Feng, 2012; Clark, 1951; Batty and Longley, 1994; Wang and Zhou, 1999). The exponential function can be expressed as follows (Takayasu, 1990)

$$\rho(r) = \rho_0 e^{-r/r_0}, \tag{9}$$

where $\rho(r)$ refers to the density at distance $r$ from the center of the city ($r=0$), $\rho_0$ to the proportionality coefficient, which is expected to be the central density, i.e., $\rho_0=\rho(0)$, and $r_0$ to the scale parameter indicating the characteristic radius of urban population distribution. Integrating equation (1) over $r$ by parts yield a cumulative distribution:

$$P(r) = 2\pi \int_0^r x\rho(x)\mathrm{d}x = P_0[1-(1+\frac{r}{r_0})e^{-r/r_0}], \tag{10}$$

in which $x$ is a distance ranging from 0 to $r$, $P(r)$ denotes the cumulative population within a radius of $r$ of the city center. The constant is derived as below

$$P_0 = 2\pi\rho_0 r_0^2, \tag{11}$$

which denotes the total population of a city. According to equation (10), the average distribution of the exponential distribution defined by equation (9) can be expressed in the following form

$$F(r) = \frac{P(r)}{A(r)} = \frac{2\rho_0 r_0^2}{r^2}\left[1-(1+\frac{r}{r_0})e^{-r/r_0}\right], \tag{12}$$

where $A(r)=\pi r^2$, and $F(r)$ denotes an average distribution of the exponential distribution defined in a 2-dimensin space. It can be proved that equation (12) bears a latent scaling. The scaling relation is as follows

$$F(\lambda r) = \frac{2r_0^2}{(\lambda r)^2}\rho_0\left[1-(1+\frac{\lambda r}{r_0})e^{-\lambda r/r_0}\right] = \lambda^{-2}\frac{2r_0^2}{r^2}\rho_0\left[1-(1+\frac{r}{r_0})e^{-r/r_0}\right] = \lambda^{-2}F(r). \tag{13}$$

In fact, if $r>r_0$ and the $r_0$ value is small enough, the cumulative distribution of urban population density satisfies the following approximate relation

$$1-(1+\frac{\lambda r}{r_0})e^{-\lambda r/r_0} \xrightarrow{r>r_0} 1-(1+\frac{r}{r_0})e^{-r/r_0}. \tag{14}$$

The solution to the functional equation (13) is an inverse power function such as

$$F(r) \propto r^{-2}, \tag{15}$$

which is just the mathematical form of the cumulative power-law distribution. The population of the central area of a city approximately satisfies this distribution (Chen and Feng, 2012).

# 3 Mathematical experiments

## 3.1 Mathematical experiment for distribution transform in 1-D space

The method of mathematical experiments can be regarded as special type of numerical experiments, which denote a process of calculations with numerical models. This process bears an analogy to classical laboratory experiments (Bowman *et al*, 1993). A mathematical experiment is defined as an approach to using computational methods based on certain mathematical models or theoretical postulates to verify a law, a relation, or an inference (Chen, 2012). In order to implement experiments, the mathematical expressions based on continuous variables must be replaced by discrete variables. Equation (5) is actually as follows

$$F(r+\delta r) = \frac{S_0}{r+\delta r}\left[1 - e^{-(r+\delta r)/r_0}\right], \qquad (16)$$

where $\delta r$ denotes a so small quantity that it can be neglected when $r$ represents a continuous variable. However, if the distance variable $r$ is discretized, we will have

$$\delta r \to \Delta r = r - (r-1) = 1.$$

Thus equation (16) changes to the following form

$$F(r+1) = \frac{S_0}{r+1}\left[1 - e^{-(r+1)/r_0}\right] \propto (r+1)^{-1}. \qquad (17)$$

Substituting $k$ for $r+1$ in equation (17) yields

$$F_k = \frac{S_0}{k}\left[1 - e^{-k/k_0}\right] \to ak^{-b}, \qquad (18)$$

where $a$ refers to proportionality coefficient, $k_0=r_0$ to the scale parameter (for the discrete format), and $b$ to the scaling exponent, which is theoretically expected to equal 1. In empirical work, the $b$ value will be close to 1.

Now, numerical experiments can be easily made by MS Excel. Suppose that the discrete form of equation (1) is $y_k=y_0*\exp(-k/k_0)$, the initial value is $y_0=100$ and the independent variable is taken as $k=0, 1, 2, \ldots, 2000$. For given scale parameter $k_0$, we can produce a geometric sequence. For example, taking $r_0=2$, we have $y_k=100, 60.6531, 36.7879, 22.3130, \ldots, 0.000$. The average sequence can be calculated as follows:

$F_1=y_0/1=100,$

$F_2=(y_0+y_1)/2=(100+60.6531)/2=80.3265,$

$F_3=(y_0+y_1+y_2)/3=(100+60.6531+36.7879)/3=65.8137,$

$F_4=(y_0+y_1+y_2+y_3)/4=(100+60.6531+36.7879+22.3130)/4=54.9385,$

......

Generally, we have

$$F_k = \frac{y_0 + y_1 + \cdots + y_k}{k} = \frac{1}{k}\sum_{r=0}^{k} y_r. \tag{19}$$

The average sequence is significantly different from the exponential sequence (Figure 1). Change $k_0$ value, we will have different geometric sequence. If $k_0$ value becomes small enough, the average sequence will approach to a straight line on the log-log plot (Figure 2). Fitting the average sequence into equation (18) yields a power law relation. For example, for $k_0=2$, we have

$$F(k) = 243.8296 k^{-0.9939},$$

The goodness of fit is about $R^2=0.9994$. The scaling exponent $b=0.9939$ is very close to 1. The smaller the $k_0$ value is, the closer the scaling exponent $b$ is to 1 (Table 1).

Table 1 The relationship between scaling exponent, goodness of fit of a 1-dimension average distribution and scale parameter of the 1-dimension exponential distribution

| $k_0$ | $S_0$ | $a$ | $b$ | $R^2$ |
|---|---|---|---|---|
| 32 | 3250.2604 | 1603.2157 | 0.8968 | 0.9751 |
| 16 | 1650.5208 | 1122.1849 | 0.9434 | 0.9882 |
| 8 | 851.0414 | 697.0211 | 0.9706 | 0.9949 |
| 4 | 452.0812 | 410.6501 | 0.9858 | 0.9981 |
| 2 | 254.1494 | 243.8260 | 0.9939 | 0.9994 |
| 1 | 158.1977 | 155.8649 | 0.9978 | 0.9999 |
| 1/2 | 115.6518 | 115.2211 | 0.9994 | 1.0000 |
| 1/4 | 101.8657 | 101.8223 | 0.9999 | 1.0000 |
| 1/8 | 100.0336 | 100.0328 | 1.0000 | 1.0000 |

**Note:** The variable $r$ and parameter $r_0$ are for continuous distributions, while the variable $k$ and parameter $k_0$ are for discrete distributions. In a mathematical experiment, $r$ and $r_0$ should be replaced with $k$ and $k_0$.

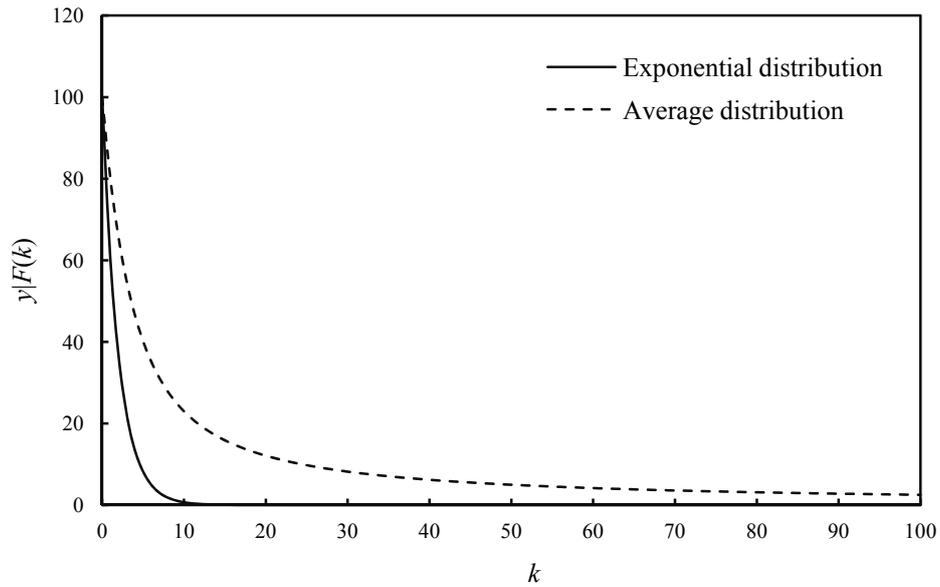

**Figure 1 An exponential distribution curve and the average distribution curve based on the 1-dimension exponential distribution**

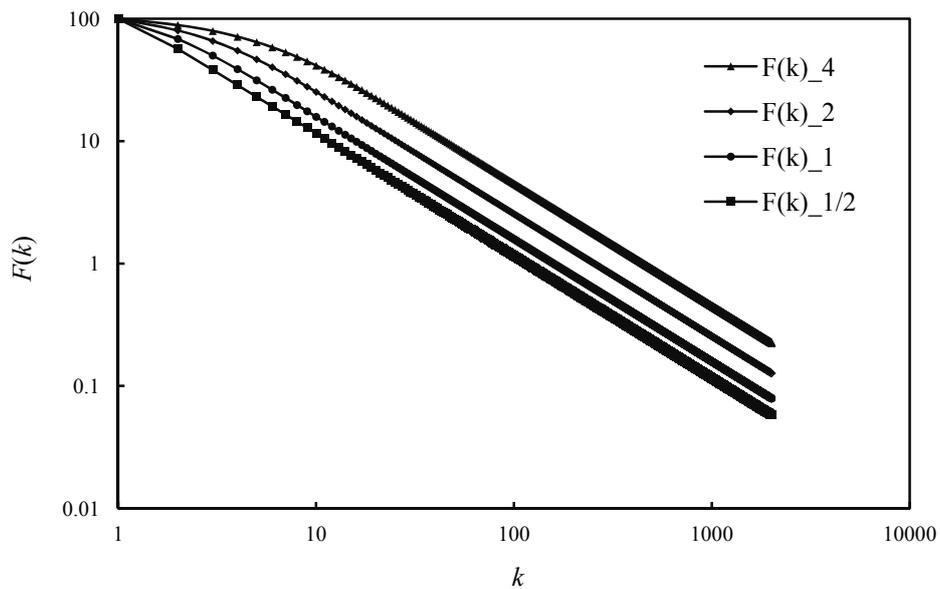

**Figure 2 Four average distribution curves based on the 1-dimension exponential distributions with scale parameter $k_0$=1/2, 1, 2, 4**

It is necessary to illustrate the scaling invariance of the 1-dimension cumulative exponential distribution, equation (2), by means of a plot. The key of the 1-dimension analysis is to demonstrate the following relation:

$$S(\lambda r) = S_0(1 - e^{-\lambda r/r_0}) \to S_0(1 - e^{-r/r_0}) = S(r), \tag{20}$$

This relation can be easily verified through numerical experiment in Excel (Figure 3). Taking $k$=1, 2, 3, … and $\lambda$=…1/2,1,2,3…, we can produce the sequence $S(k)$ and $S(\lambda k)$. In fact, if $k_0$ value is small, and $k$ value is large, there will be no significant difference between $S(\lambda k)$ and $S(k)$. That is, $S(\lambda k) \to S(k)$ or $S(\lambda k) \approx S(k)$ for small $k_0$ and large $k$.

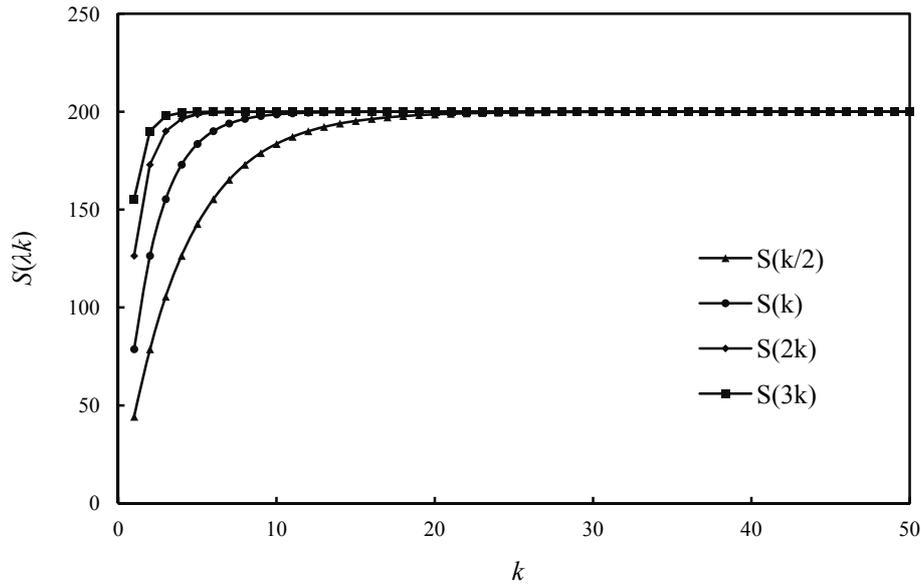

**Figure 3 A pattern of scaling invariance of the 1-dimensin cumulative exponential distribution**

**($k_0$=2, $\lambda$=1, $\lambda$=2, $\lambda$=3)**

**3.2 Mathematical experiment for distribution transform in 2-D space**

In practice, the 2-dimension-based numerical experiment is more difficult than the 1-dimension-based numerical experiment. Suppose that the discrete form of equation (9) is $\rho_k = \rho_0 * \exp(-k/k_0)$, the initial value is taken as $\rho_0$=100 and the variable is also $k$=0, 1, 2, …, 2000. For given scale parameter $k_0$, we can create a series of exponential decay. For example, taking $k_0$=2, we have a density sequence such as $\rho_k$=100, 60.6531, 36.7879, 22.3130, … , 0.000. Correspondingly, the area sequence is: $A_k = \pi k^2$=0, 3.1416, 12.5664, 28.2743,…, 12566370.614, and the sequence of ring area is as below: $\Delta A_k = \Delta(\pi k^2) = \pi\Delta[k^2-(k-1)^2)]$=0, 3.1416, 9.4248, 15.7080,…,12563.2290 (for $k$=0, let $\Delta A_0$=0). The average sequence can be calculated with the following formula

$$F_k = \frac{\rho_0 \Delta A_0 + \rho_1 \Delta A_1 + \cdots + \rho_k \Delta A_k}{A_k} = \begin{cases} \rho_0, & k = 0 \\ \dfrac{\rho_0 + \pi \sum_{i=1}^{k} \rho_i [i^2 - (i-1)^2]}{\pi k^2}, & k > 0 \end{cases} \quad (21)$$

The first four data points are as follows:

$F_0 = \rho_0 = 100$,

$F_1 = [\rho_0 + \rho_1 \pi (1^2 - 0^2)]/(\pi 1^2) = (100 + 60.6531*3.1416*1)/3.1416 = 92.4841$,

$F_2 = [\rho_0 + \rho_1 \pi (1^2 - 0^2) + \rho_2 \pi (2^2 - 1^2)]/(\pi 2^2)$

$\quad = (100 + 60.6531*3.1416*1 + 36.7879*3.1416*3)/3.1416*4 = 50.7120$,

$F_3 = [\rho_0 + \rho_1 \pi (1^2 - 0^2) + \rho_2 \pi (2^2 - 1^2) + \rho_3 \pi (3^2 - 2^2)]/(\pi 3^2)$

$\quad = (100 + 60.6531*3.1416*1 + 36.7879*3.1416*3 + 22.3130*3.1416*5)/3.1416*9 = 50.7120$,

......

The intuitional difference between the exponential distribution and the average exponential distribution is not significant on diagram (Figure 4). Change $k_0$ value, we will have different exponential sequence. If $k_0$ value becomes small enough, the average sequence will evolve into a straight line on the log-log plot (Figure 5). Fitting the average sequence into equation (15) yields a power law model. For instance, for $k_0=2$, we have

$$F(k) = ak^{-b} = 594.4477 k^{-1.9843}.$$

The determination coefficient is around $R^2=0.9993$. The scaling exponent $b=1.9843$ is close to 2. The smaller the $k_0$ value is, the closer the scaling exponent $b$ is to 2 (Table 2).

Table 2 The relationship between scaling exponent, goodness of fit of a 2-dimension average distribution and scale parameter of the 2-dimension exponential distribution

| $r_0$ | $P_0$ | $a$ | $b$ | $R^2$ |
|---|---|---|---|---|
| 32 | 633548.9832 | 32443.7015 | 1.7337 | 0.9658 |
| 16 | 156026.0895 | 18056.1034 | 1.8518 | 0.9834 |
| 8 | 37900.6008 | 7102.6184 | 1.9221 | 0.9928 |
| 4 | 8994.8042 | 2217.0778 | 1.9623 | 0.9973 |
| 2 | 2077.2877 | 594.4477 | 1.9843 | 0.9993 |
| 1 | 495.6429 | 152.8531 | 1.9953 | 0.9999 |
| 1/2 | 164.5639 | 52.1738 | 1.9994 | 1.0000 |
| 1/4 | 106.0801 | 33.7639 | 2.0000 | 1.0000 |
| 1/8 | 100.1055 | 31.8646 | 2.0000 | 1.0000 |

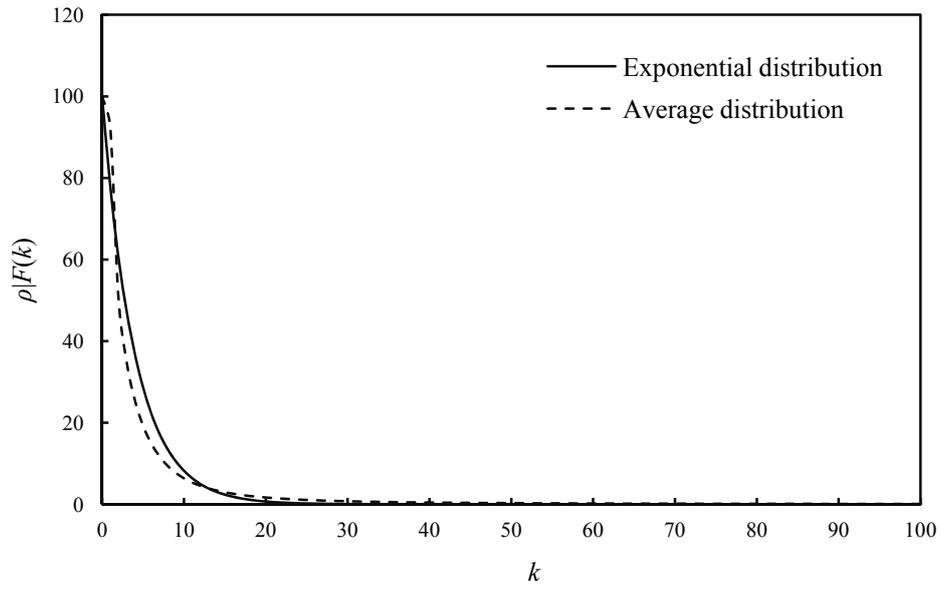

**Figure 4 An exponential distribution curve and the average distribution curve based on the 2-dimension exponential distribution**

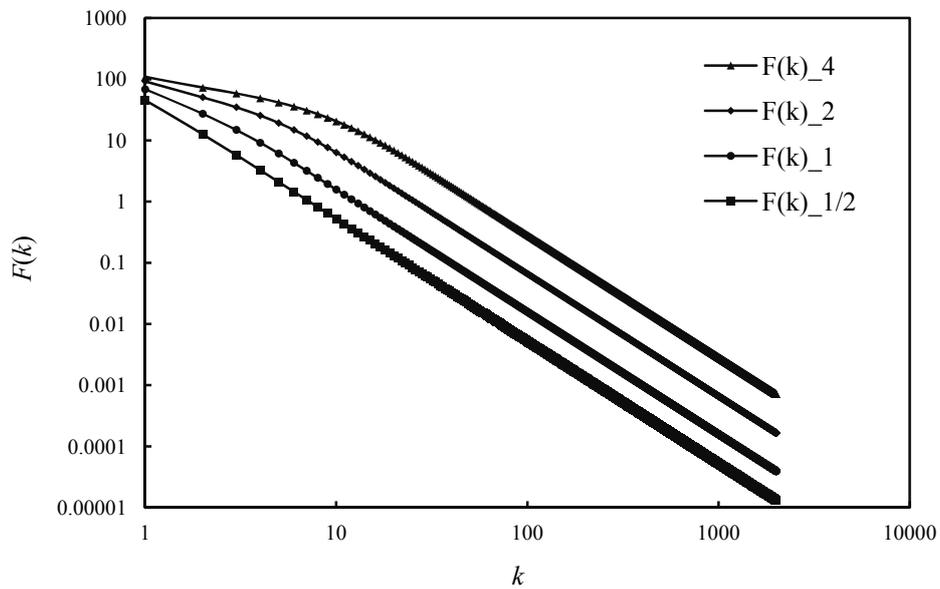

**Figure 5 Four average distribution curves based on the 2-dimension exponential distributions with scale parameter $k_0$=1/2, 1, 2, 4**

The scaling invariance of the 2-dimension cumulative exponential distribution, equation (10), can also be illustrated with a plot. The key of the 2-dimension analysis is to demonstrate the following relation:

$$P(\lambda r) = P_0[1-(1+\frac{\lambda r}{r_0})e^{-\lambda r/r_0}] \to P_0[1-(1+\frac{r}{r_0})e^{-r/r_0}] = P(r), \qquad (22)$$

This relation can be visually verified with numerical experiment (Figure 6). It is easy to create the series of $P(k)$ and $P(\lambda k)$ by taking $k=1,2,3,...$ and $\lambda=...1/2,1,2,3...$. If the $k_0$ value is small, and $k$ value is large, there will be no significant difference between $P(\lambda k)$ and $P(k)$. That is, $P(\lambda k) \to P(k)$ or $P(\lambda k) \approx P(k)$ for small $k_0$ and large $k$.

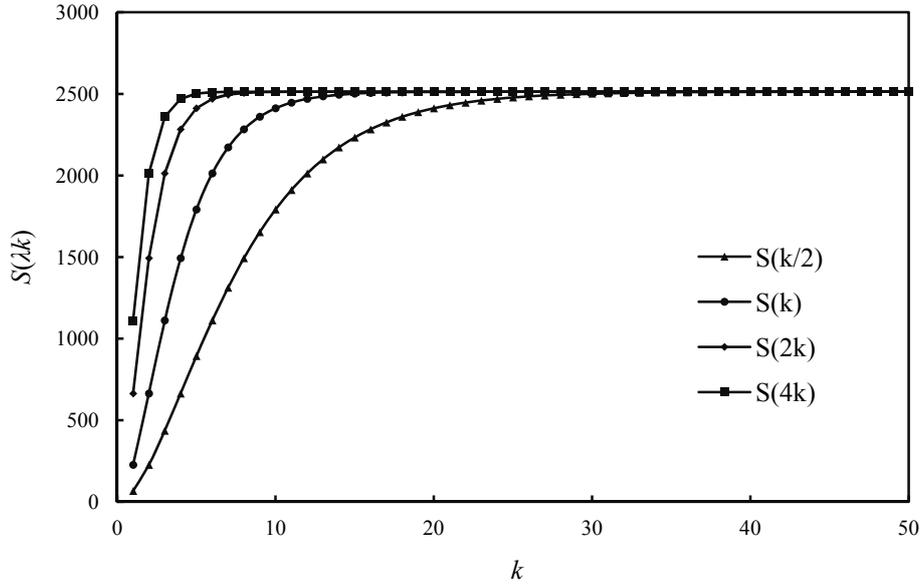

**Figure 6 A pattern of scaling invariance of the 2-dimension cumulative exponential distribution**

**($k_0=2, \lambda=1, \lambda=2, \lambda=3$)**

## 3.3 Supplementary explanation

The mathematical experiment of the 2-dimension exponential distribution is similar to that of the 1-dimension exponential distribution, but there is subtle difference. For the 1-dimension exponential distribution, the process of averaging the density numbers leads to 1-step displacement of time or space variable $r$, and $r$ should be replaced with $k=r+1$ for the average series. However, for the 2-dimension exponential distribution, the averaging process does not result in variable displacement. Therefore, it is unnecessary to rescale the independent variable. If the spatial distribution is isotropic, the 2-dimension exponential distribution can be defined in 1-dimension space so that the mathematical process becomes simple. However, if we calculate the cumulative distribution or average distribution, the data processing must be based on a 2-dimension space.

No matter for which exponential distribution, 1-dimension or 2-dimension, we have two approaches to creating a sample path for the cumulative distribution or average distribution. One approach is to sampling in the discrete way, and the other, is to select discrete data points in a continuous way. For the 1-dimension exponential distribution, if we use equation (19) to calculate the sample path of average exponential distribution, the way and result are discrete; if we use equation (5) to generate a sample path by substituting $k$ for $r$ ($r$=0, 1, 2, …; $k=r+1$=1,2,3,…), the result is discrete but the way is continuous. For the 2-dimension exponential distribution, if we use equation (21) to obtain the sample path of average distribution, the way and result are discrete; if we use equation (12) to create a sample path, the result is a discrete form but the method is a continuous way. The sampling result from the discrete way is different from that from the continuous way, but they are equivalent to one another. The ratio of the data point from a discrete sampling to that from a continuous sampling is a constant. For example, for the 1-dimension average exponential distribution with a scale parameter $r_0$=1, if the variable $k=r+1$ ranges from 1 to 2001, the scaling exponent is about $b$=0.9978, and the goodness of fit is around $R^2$=0.9999. However, the discrete sampling result gives a proportionality coefficient $a$=155.8657, while the continuous sampling result gives a coefficient $a$=98.5259 (Figure 7). The ratio of the latter to the former is about 0.6321.

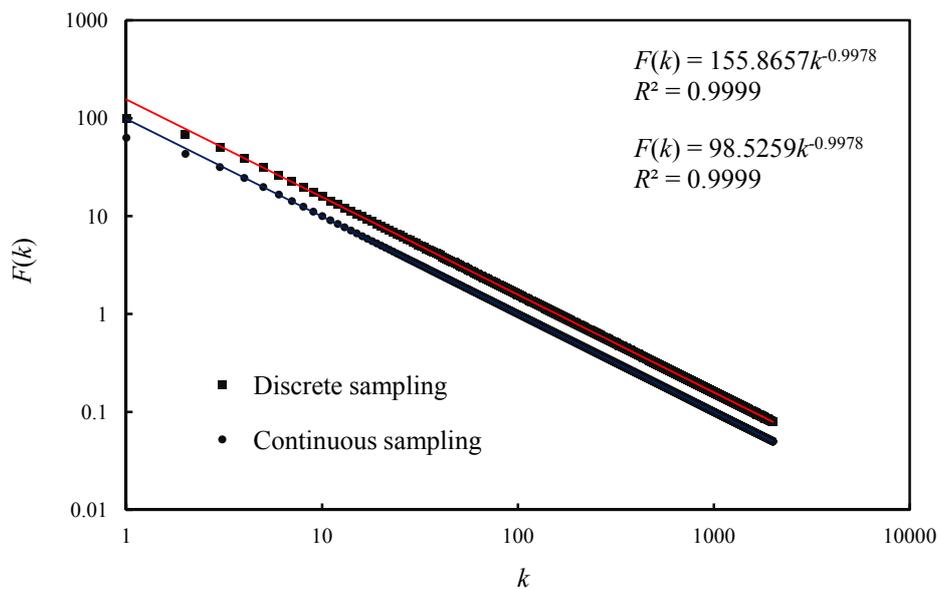

Figure 7 The 1-dimension average distributions based on continuous sampling and discrete

# 4 Theoretical interpretation

## 4.1 Comparison between exponential law and power law

The exponential distribution is significantly different from the power-law distribution despite the similar decay curves of profiles. Batty and Kim (1992) once discussed the similarities and differences between power laws and exponential laws. A comparison between exponential laws and power laws can be drawn as below (Table 3). The power-law distribution has no characteristic scale (length), and it suggests a scaling process. Oppositely, the exponential distribution has a characteristic scale (length). The scale parameter, or the reciprocal of the rate parameter, indicates a typical length which is associated with the average value. For the 1-dimension exponential distribution, the characteristic scale parameter is as follows

$$r_0 = \frac{S_0}{y_0}; \tag{23}$$

For the 2-dimension exponential distribution, the characteristic scale parameter is as below

$$r_0 = \sqrt{\frac{P_0}{2\pi\rho_0}}. \tag{24}$$

Obviously, for given total quantity $S_0$ or $P_0$ and initial value $y_0$ or $\rho_0$, the parameter $r_0$ value is certain. Using the method of entropy-maximization, we can demonstrate that the scale parameter represents some kind of mean value (Chen, 2008; Chen, 2012).

Table 3 Similarities and differences between exponential distributions and power-law distributions

| *Item* | *Exponential distributions* | *Power-law distributions* |
| --- | --- | --- |
| Curve form | Quasi-heavy-tailed distribution | Heavy-tailed distribution |
| Process | Scale translation | Scaling transform |
| Scale | Typical scale | No typical scale |
| Scaling | Linear scaling | Nonlinear scaling |
| Correlation function | Exponential distribution | Power-law distribution |
| Auto-correlation | Tail off | Tail off (heavy tail) |
| Partial auto-correlation | Cut off (1 order) | Tail off (light tail) |
| Memory | No memory | Long (term/distance) memory |
| Dynamics | Simplicity | Complexity |
| Symmetry | Translational invariance | Scaling invariance |
| Physical basis | Entropy-maximization (single process) | Entropy-maximization (dual process) |

The average value (mean) is basic and very important for statistical analysis. It is a position parameter that indicates the mass center of data points. Based on the *mean*, we can calculate *variance*, a squared distance parameter indicative of differentiation, and *covariance*, an inclination parameter indicative of correlation in statistical analysis. If the values of mean, variance, and covariance are determined, the statistical structure of a pattern or process of a system or its evolution is predictable, and thus the system is a type of simple system. Inversely, if we cannot find a valid mean for a distribution, we will be unable to know the statistical structure of a system, and thus cannot predict its trend of development. In this case, the system is complex. Therefore, the exponential distribution indicates simple systems, while the power-law distribution suggests complex systems (Barabási, 2002; Chen and Zhou, 2008; Goldenfeld and Kadanoff, 1999). A comparison between simple systems and complex systems can be drawn as follows (Table 4). It is unexpected that the power-law distribution can be created from the average process of an exponential distribution. This finding suggests a new way of looking at complexity from the angle of view of simplicity.

**Table 4 Comparison between simple systems and complex systems**

| *Item* | *Simple system* | *Complex system* |
|---|---|---|
| Methodology | Reductionism is valid | Reductionism is invalid |
| Element relationships | Linear relationships | Nonlinear relationships |
| Probability distribution | With characteristic scale (mean is valid) | No characteristic scale (mean is invalid) |
| Interaction | No spatio-temporal lag effect | With spatio-temporal lag effect |
| Law | Symmetry | Asymmetry |
| Prediction | Predictable | Unpredictable |

## 4.2 A link between simplicity and complexity

In literature, the power law distribution is always associated with scaling, and the scaling process is always associated with complexity. In fact, both exponential and power law distributions can be related to the concept of scaling. However, the exponential distribution indicates a linear scaling, while the power distribution suggests a nonlinear scaling. The linear scaling implies simplicity. Complexity can be only associated with nonlinear scaling. For example, substituting a discrete variable $m$ for the continuous variable $r$, we can discretize equation (1) and get a new expression as below

$$y_m = y_0 e^{-1/r_0}(e^{1/r_0})^{1-m} = y_1 r_y^{1-m}, \qquad (25)$$

where $y_1 = y_0 \exp(-1/r_0)$ refers to a constant coefficient, and $r_y = \exp(1/r_0)$ to a common ratio, $m$ denotes a sequence of natural numbers ($m=1,2,3,\ldots$). A proportional relation can be derived from equation (25) such as

$$y_m = r_y y_{m+1}. \qquad (26)$$

Apparently, this is a type of linear scaling (Williams, 1997).

Two linear scaling relations support a nonlinear scaling relation. Suppose that there is another exponential equation in the form

$$x_m = x_1 r_x^{\pm(1-m)}, \qquad (27)$$

which is parallel to equation (25). Then a power law can be derived from equations (25) and (27), and the result is as follows

$$y_m = \mu x_m^{\pm\sigma}, \qquad (28)$$

where $\mu = y_1 x_1^{\pm\sigma}$, $\sigma = \ln r_y / \ln r_x$. Equation (28) suggests a nonlinear scaling relation

$$f(\lambda x_m) = \mu(\lambda x_m)^{\pm\sigma} = \lambda^{\pm\sigma} f(x_m), \qquad (29)$$

in which $\mu$ denotes a proportionality coefficient, and $\sigma$ is a scaling exponent. The mathematical deduction from equation (25) to equation (28) means that the relationship between an exponential function and a power function indicates a link between simplicity and complexity. This paper lends further support to this judgment. In urban studies, equation (28) can be used to represent allometric growth, Pareto distribution, and Zipf's law (Newman, 2005). The power function can be decomposed into two exponential functions such as equations (25) and (27) (Chen, 2012; Chen and Zhou, 2003; Jiang and Yao, 2010).

## 4.3 A spurious Zipf distribution and fictitious fractals

The 1-dimension exponential distribution can result in fake Zipf distribution. Zipf's law is frequently observed within the natural world as well as in human society (Bak, 1996; Zipf, 1949). If the elements of a system take on a rank-size distribution, they are regarded as following Zipf's law. One of the most conspicuous empirical facts in the social sciences is Zipf's distribution of cities (Gabaix, 1999a; Gabaix, 1999b; Jiang and Jia, 2011; Rozenfeld *et al*, 2011). The 'law' states that, if the size of a city is multiplied by its rank, the product will equal the size of the highest ranked city. In other words, the size of a city ranked $k$ will be $1/k$th of the size of the largest city. The mathematical expression of Zipf's law is

$$Z_k = Z_1 k^{-1}, \tag{30}$$

where $Z_k$ refers to the size of the city ranked $k$, and $Z_1$ to the size of the largest city. Apparently, equation (8) and equation (30) bear similar function form. In fact, if the scale parameter $r_0$ value is small enough (say, $r_0 < 0.05$), the 1-dimension exponential function equation (1) will reduce to a step function

$$y = \begin{cases} y_0, & r = 0 \\ 0, & r > 0 \end{cases}. \tag{31}$$

The average distribution of the step distribution is just the Zipf's distribution. When the scale parameter value becomes smaller and smaller, the average distribution of the exponential distribution will become closer and closer to Zipf's distribution. The Zipf distribution is just the extreme case of the average exponential distribution. However, when the scale parameter is not small, the scaling exponent of the average exponential distribution is very close to 1 (Figure 8). This kind of power-law distribution can be termed spurious Zipf distribution.

The spurious Zipf distribution can be reduced to an exponential distribution. The reduction formula is as follows

$$y_{k-1} = kF_k - (k-1)F_{k-1}, \tag{32}$$

in which $F_k$ represents the spurious Zipf sequence, while $y_{k-1}$ denotes an geometric sequence

($k=1,2,3,\ldots$). For example, for the first data points, the reduction formula can be expressed as below

$$y_0 = F_1$$
$$y_1 = 2F_2 - y_0 = 2F_2 - F_1$$
$$y_2 = 3F_3 - y_0 - y_1 = 3F_3 - 2F_2$$

Using these formulae and repeating computation, we can reduce a power-law distribution to an exponential distribution.

Suppose there is rank-size distribution following a power law. If the power-law distribution can be reduced to an exponential distribution, it is a spurious Zipf distribution; if the distribution can be reduced to a step distribution, it is a standard Zipf distribution; if it cannot be reduced to an exponential distribution, it is a general Zipf distribution. These results provide a new way of looking at the fake Zipf distribution. There are various spurious Zipf distributions. The spurious "rank-size patterns" remind us of the length distribution of the words produced by the well-known monkey of Miller (1957), who postulated that the probability of a word of length $i$ generated by this monkey will decrease exponentially as $i$ increases ($i=1, 2, 3, \ldots$).

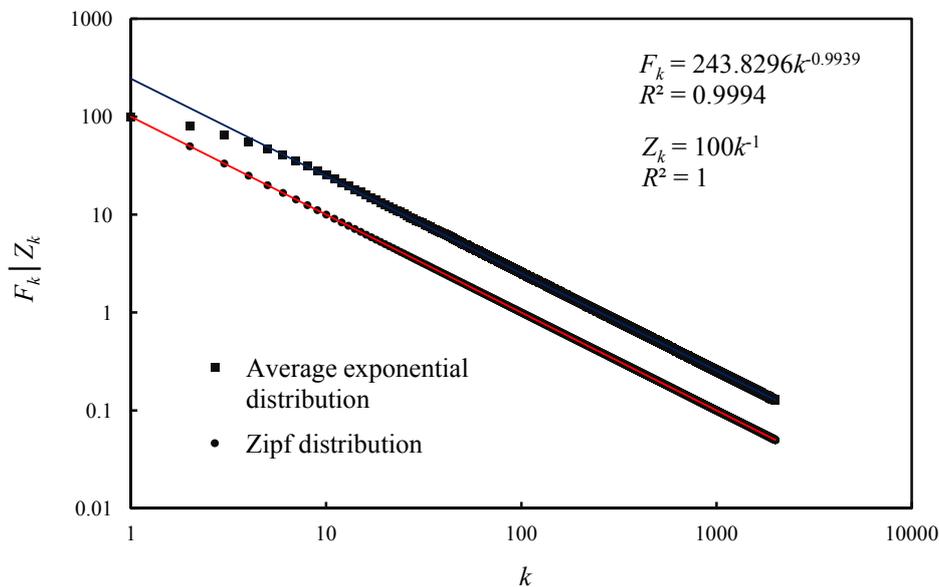

**Figure 8 The 1-dimension average exponential distribution ($k_0=2$) and the Zipf distribution**

**Note:** The above points and line indicate the spurious Zipf distribution, which come from an average exponential distribution, while the below points and line represent a real standard Zipf distribution.

The 2-dimension exponential distribution can lead to a fictitious fractal dimension. Fictitious fractals are sometimes called 'fractal rabbits'. Kaye (1989, page 24) pointed out: 'This term was derived from the fact that the false fractal appears out of nowhere like a white rabbit out of a magician's hat.' In special cases, an exponential distribution can be associated with self-affine fractals (Chen and Feng, 2012). However, generally speaking, the exponential distribution does not indicate fractal structure or form. Despite this fact, a 'fractal rabbit' may 'jump' out of an exponential distribution by average processing or Fourier transform. In Table 2, some scaling exponent values can be confused with fractal dimension, but they are not real fractional dimension.

# 5 Conclusions

This is a paper concerning exponential distributions, but it proposes a new approach to understanding the power-law distribution. Though a power law differs from an exponential law significantly, the links between power-law distributions and exponential distributions can be found. Exponential law indicates simplicity, while power law suggests complexity. The mathematical connections between power laws and exponential laws are helpful for our exploring complex systems. Now, the main conclusions of this research can be drawn as follows.

**First, power-law distributions can be derived from exponential distribution by average processing.** The cumulative distribution of an exponential distribution possesses a hidden scaling. If we fix the initial data point to make a moving average of an exponential distribution, we will obtain a power-law distribution. If the exponential distribution is define in 1-dimension space, the scaling exponent of the power-law distribution will be close to 1; if the exponential distribution is defined in 2-dimension space, the scaling exponent of the power-law distribution will approach to 2. The scaling exponent value suggests the spatial dimension of exponential distributions.

**Second, the power-law distributions based on exponential distributions are different from the general power-law distributions.** The moving average of an exponential distribution is very similar in form to a power-law distribution. However, there is a subtle or even an essential difference between them. A real power-law distribution cannot be reduced to an exponential distribution, but the power-law distributions derived from exponential distributions can be reduced to exponential distributions. The mathematical transformation from exponential distributions to

power-law distributions and the inverse process provide a new approach to identifying the spurious Zipf distribution and fictitious fractal dimension.

**Third, the relationships between exponential laws and power laws make a bridge between simplicity and complexity.** An exponential distribution indicates simplicity because of its characteristic scale parameter, while a power-law distribution suggests complexity because of its scale-free property. What interest scientists is the inherent connection between simplicity and complexity. Why simple interactions in a determinate system can lead to complex behaviors such as bifurcation and chaos? What are the simple rules which dominate the dynamic behavior of complex systems? The studies on the link between exponential laws and power laws may suggest a new way of answering these pending questions.

## Acknowledgment

This research was sponsored by the National Natural Science Foundation of China (Grant No. 41171129). The supports are gratefully acknowledged.